\documentstyle[12pt]{article}
\begin{document}
\title{New meson mass relation and lowest 
pseudoscalar glueball mass
\thanks{This work is supported in part by 1) NSFC under
contract No. 19677102, No. 19775044 and No. 19991480; 
2) DURF of SEC under contract No. 97035807; 3) BEPC National 
Lab of China; 4) CTNP of LHIA National Lab of china and 
5) China Postdoctoral Science Foundation. }}

\vskip 0.8in

\author{{Ning Wu$^1$, Tu-Nan Ruan$^{2,3}$, Zhi-Peng Zheng$^{1,2}$  }\\
{\small Division 1, Institute of High Energy Physics, P.O.Box 918-1,
Beijing 100039, P.R.China$^1$}  \\  
{\small CCAST(World Lab), Beijing  100080, P.R.China $^2$ }\\
{\small Dept. of Modern Phys., Univ. of Sci. and 
Tech. of China, Hefei, Anhui 230026, P.R.China $^3$  }  \\
}

\maketitle
\vskip 0.8in

\vskip 0.3in

\noindent
\begin{abstract}

After considering mixing with glueball, we give a new
mass relation for meson nonet. According to this mass relation and
predicted mass of pseudoscalar glueball given by
lattice calculation and effective Hamiltonian, 
the expected mass of mixed pseudoscalar glueball  
is about 1.7 Gev. This result  is helpful for
experimental search for the mixed 
Isoscalar pseudoscalar glueball. 
$\eta(1760)$ is discussed as 
a possible canditate of this kind of particle.

\end{abstract}

\newpage

\vskip 0.3in


There is some inconsistency about meson mass spectrum and 
lowest glueball mass between theory and experiments. As an
example, the lowest mass of pseudoscalar glueball which
is given by lattice calculation is about 
2232 $\pm$ 370 $\pm$ 220 Mev \cite{1}.
However, the mass value of experimental candidate is much 
lower than this value. Another question is about 
Gell-mann-Okubo formula\cite{3}. It is know that vector
meson nonet and baryon decuplet satisfy the 
Gell-mann-Okubo formula well. But the pseudoscalar
meson nonet does not satisfy the Gell-mann-Okubo formula 
at all. Why pseudoscalar meson nonet does not satisfy it?
The third question is about glueball-meson mixing. 
It is believed that there is mixing between meson and
glueball. If there is mixing, how to determine the masses
of glueball and mesons after mixing? 
A meson mass relation can 
be obtained after consider glueball-meson mixing. This 
mass relation is almost well satisfied by all 
meson nonet\cite{9}, and it can help us to determine 
the mass of some mixed glueball.\\

First, let's discuss the general case. Suppose that there is
a mason nonet. $M^0_1$ and $M^0_8$ are $SU(3)$ 
singlet states and the isoscalar of $SU(3)$ nonet states. 
The nearest glueball which has the same $I^G J^{PC}$ as
the nonet is denoted as $G^0$. Generally, these three 
states will have mixing. The corresponding states after
mixing are denoted as $M_1$, $M_8$ and $G$. Suppose
that the mixing matrix is $U$, then we have
$$
\left (
\begin{array}{c}
M_1  \\
M_8   \\
G
\end{array}
\right )
= U
\left (
\begin{array}{c}
M^0_1  \\
M^0_8   \\
G^0
\end{array}
\right ).
\eqno{(1)}
$$
In the most general case, the mixing matrix is a $3 \times 3$
unitary matrix. However, in the present case, it is a real
orthogonal matrix. If the mass matrix before mixing is denoted
as $M^0$, then the mass matrix after mixing is given by:
$$
M =  U M^0 U^t,
\eqno{(2)}
$$
where $U^t = U^{-1}$ is the transposition matrix of $U$. The 
diagonal elements of mass matrix $M^0$ are $m_{M^0_1}$,
$m_{M^0_8}$ and $m_{G^0}$ respectively, and the 
diagonal elements of mass matrix $M$ are $m_{M_1}$,
$m_{M_8}$ and $m_{G}$ respectively.  
The trace of eq.(2) gives the following relation:
$$
  m_{M^0_1} + m_{M^0_8} + m_{G^0}
= m_{M_1}   + m_{M_8}   + m_{G}.
\eqno{(3)}
$$
The particles that have isospin 1 in the meson nonet are 
denoted as $M_{\pi}$ and the particles have isospin 
$\frac{1}{2}$ are denoted as $M_K$. Supposed that
Gell-Mann-Okubo mass formula are well satisfied by
meson nonet before mixing, that is
$$
4 m_{M_K} - m_{M_{\pi}} = 3 m_{M^0_8}
\eqno{(4)}
$$
$$
2 m_{M_K} + m_{M_{\pi}} = 3 m_{M^0_1}.
\eqno{(5)}
$$
Then we have
$$
2 m_{M_K} = m_{M^0_8} + m_{M^0_1}.
\eqno{(4)}
$$
Then eq(3) will change into:
$$
  2 m_{M_K} + m_{G^0}
= m_{M_1}   + m_{M_8}   + m_{G}.
\eqno{(6)}
$$
This is the mass relation for meson nonet. In the
above relation, $M_1$, $M_8$ and $M_K$ are mesons which
have already been observed in experiments and their masses
are already known. $G^0$ are pure glueball, its mass are
predicted by theoretical calculations. So, in the above
relation, the only unknown thing is the mass of mixed glueball
$G$. It can be calculated from the above relation. 
Therefore, this meson mass relation can help us to 
determine the mass of mixed glueball $G$. \\

Now, let's use this mass relation to discuss the lowest
mass of pseudoscalar glueball. According to PDG\cite{2}, 
the two pseudoscalar isoscalar mesons are $\eta$ 
and $\eta'$.  Their masses are
$$
m_{\eta}= 547~ {\rm Mev},
~~~~ m_{\eta'}= 958~ {\rm Mev}. 
\eqno{(7)}
$$
The isospin $\frac{1}{2}$ mesons are Kaons, their masses
are $m_K = 494$ Mev. Then eq(6) changes into:
$$
m_{G} = m_{G^0} - 517 ({\rm Mev}).
\eqno{(8)}
$$
In the above relation, $m_{G^0}$ is the mass of pure 
pseudoscalar glueball. Its mass can be obtained from
theoretical calculation. The calculated mass of pure 
pseudoscalar glueball from lattice calculation 
is $2.23 \pm 0.37 \pm 0.22$ Mev\cite{1}. 
Then eq(8) means that the center value of the 
lowest mass of mixed  glueball is about 1713 Mev. 
However, its predicted center mass from an effective Hamiltonian
is about $2.19$ Mev. Then the lowest center mass
of mixed glueball is about $1673$ Mev. 
According to PDG, there are two pseudoscalar mesons
below 2.0 Gev, there are $\eta(1440)$ and
$\eta(1760)$. 
The mass of $\eta(1760)$ is more closer to the predicted masses
given by both lattice calculation and effective Hamiltonian.
The width of this resonance is  
narrow(only about 60 Mev)\cite{5}. This resonance is also
found in $J/\psi \to \gamma K \bar{K} \pi$ and
$J/\psi \to \gamma \eta \pi \pi$\cite{6}. 
Using all these experimental results, the partial width of
$\eta(1760)$ decays into  two gluons can be calculated\cite{7}. 
It is found that this partial width is big.
From these results,  especially its mass position, 
it seems that $\eta(1760)$ is more like a mixed glueball. 
But in order to confirm its glueball nature of $\eta(1760)$,
a lot of things need to do, especially accurately measurement
of the total branching ratio
of of $J/\psi \to \gamma \eta(1760)$, its mass and
width,  and detailed study on its decay pattern. \\

An evidence to support the glueball nature of 
$\eta(1760)$ is about the study of 
$J/\psi \to \gamma \gamma \rho^0$\cite{8}. 
For $\eta(1440)$ appears in the invariant mass spectrum
of $\gamma \rho^0$, and $\eta(1760)$ does not, and
glueball should have very weak coupling with photons. 
Another evidence to support the glueball nature of
$\eta(1760)$ is about $J/\psi$ radiative to $\gamma \eta'$. 
The mass of $\eta$ is close to the mass
of $\eta_8$, but the mass of $\eta'$ is much higher then 
the mass of $\eta_1$. The glueball content is mainly mixed
into $\eta'$ so as to increase its mass by about 500 Mev.
In another words, the meson $\eta'(958)$ should contain 
large amount of glueball content. 
This result is consistent with experimental results, because
$\eta'$ is abundantly produced in
radiative $J/\psi$ decay. The branching ratio
of $J/\psi \to \gamma \eta'$ is big\cite{2}, it is
$4.31 \times 10^{-3}$. The large sample of $J/\psi$
data in BES provides a good oppotunity for the 
comfirmation of the mixed Isoscalar pseudoscalar glueball.
\\

Eq.(6) is also helpful for the scalar glueball and tensor
glueball searching\cite{9}. 
For example, the two isoscalar tensor
mesons are $f_2(1270)$ and $f'_2(1525)$, and the isospin
$\frac{1}{2}$ tensor meson is $K_2^*(1430)$. According
to eq.(6), we have $m_{G} \approx m_{G^0}.$
It means that the mass of the mixed tensor glueball is 
very close to the mass of pure glueball. The mixing
between glueball and ordinary tensor mesons is small. 
The mass of tensor glueball should close to the 
theoretical predictions. 
\\

\end{document}